\documentclass[letterpaper, 10 pt, conference]{ieeeconf}  

\IEEEoverridecommandlockouts                              

\usepackage{cite}

\usepackage{xr}

\usepackage{amssymb}
\usepackage{graphicx} 

\usepackage{multicol}        
\usepackage{multirow}        
\usepackage{amsmath}                      
\usepackage{amsfonts}                     
\usepackage{mathrsfs}                     
\usepackage{amssymb}
\usepackage{float}
\usepackage{mathtools}
\usepackage{nicefrac}
\usepackage{subfig}
\usepackage[font=small,labelfont=bf]{caption}

\usepackage{algorithm}
\usepackage[noend]{algorithmic}

\usepackage[dvips,frame,rotate,import]{xy} 

\usepackage{url}
\usepackage{color} 
\usepackage{xspace}
\title{\LARGE \bf
Multi-Defender Single-Attacker Perimeter Defense Game on a Cylinder: Special Case in which the Attacker Starts at the Boundary
}

\author{Michael Otte and Roderich Gro{\ss}
\thanks{This work was inspired by the MRS'21 work of Guerrero-Bonilla {\it et al.}~\cite{GuerreroBonilla.etal.MRS21} --- which investigated a multi-agent perimeter patrol game.
Michael Otte and Roderich Gro{\ss} formulated a variation of this game during the MRS'21 conference held in Cambridge, UK,
being energized by the conference as well as the ability to interact with colleagues in-person again. 
In addition to describing an interesting multi-agent problem, the authors believe their paper highlights the collaborative nature of the MRS community and importance of interactions that take place at the MRS conferences.}
}

\begin{document}

\newcommand{\argmin}{\operatornamewithlimits{arg\ min}}
\newcommand{\argmax}{\operatornamewithlimits{arg\ max}}

\newcommand{\probability}[1]{\mathbb{P}(#1)}

\newcommand{\algOr}{\mathbf{or}}
\newcommand{\algAnd}{\mathbf{and}}

\newcommand{\bigOh}[1]{\mathcal{O}(#1)}
\newcommand{\bigTheta}[1]{\Theta(#1)}
\newcommand{\bigOmega}[1]{\Omega(#1)}

\newcommand{\Cicumfrence}{C}
\newcommand{\dist}{d}
\newcommand{\velocity}{v}

\newcommand{\pose}{x}
\newcommand{\adversary}{a}

\newcommand{\agent}{\mathrm{agent}}
\newcommand{\gap}{\mathrm{gap}}

\newcommand{\numAgents}{n}

\newcommand{\agentIndex}{i}
\newcommand{\agentIndexB}{j}

\newcommand{\timeVar}{t}

\newcommand{\param}{\gamma}

\maketitle
\thispagestyle{empty}
\pagestyle{empty}

\begin{abstract}
We describe a multi-agent perimeter defense game played on a cylinder. A team of $n$ slow-moving defenders must prevent a single fast-moving attacker from crossing the boundary of a defensive perimeter. We describe the conditions necessary for the attacker to win in the special case that the intruder starts close to the boundary and in a region that is currently defended, and assuming that the defenders are optimally positioned. 
\end{abstract}

\section{Introduction}

The defense of a perimeter/boundary is a scenario that arises across a variety of domains, for example, securing property, preventing poaching in wildlife refuges, providing military defense, etc. In some settings it is reasonable to assume that there are multiple defensive agents that guard the perimeter, and a single offensive agent that attempts to infiltrate the perimeter. The defensive agents are called `defenders' and the offensive agent is called the `attacker.'

\begin{figure}[b!]

\includegraphics[width=3.4in, trim=0 0 0 2, clip=true]{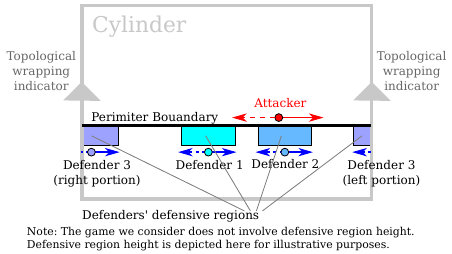}

  \caption[]{In this work we consider a special case of a multi-agent perimeter defense game in which the defense boundary exists in a topological cylinder and an attacker starts close to the perimeter boundary. The attacker moves more quickly than the defenders --- maximum speed of the attacker and defenders is indicated by the red and blue arrow lengths. Note that the cylinder wraps around such that the left and right sides of the environment are connected (as illustrated by the depiction of Defender 3).}
  \label{fig:basicSetup}
\end{figure}

We consider 
a  ``multi-defender single-attacker perimeter defense game'' where
multiple defenders defend a continuous boundary within a (topological) cylinder (see Figure~\ref{fig:basicSetup}). Each defender prevents the attacker from crossing the perimeter's boundary within a small region centered around the defender. Defenders move along the boundary at constant speed and can change direction instantaneously. The defenders are homogeneous in the sense that they have identically sized defensive regions and traveling velocities. The defenders move more slowly than the attacker, hence, they must work together to guard the boundary.

In this work, we consider the special case of the aforementioned game in which the attacker starts infinitesimally close to the boundary in a region that is currently defended. We assume that all agents (defenders and attacker) are aware of the locations of all other agents. Given these assumptions, we derive closed-form expressions for the cases where the attacker can win even when the defenders start in an optimal configuration and play an optimal strategy. The expressions that we derive involve the perimeter length, the attacker's velocity, the number of defenders, the defenders' velocities and defense region sizes.

\section{Motivation of Scenario and Special Case}

\begin{figure}[b!]

\includegraphics[width=3.4in, trim=0 0 0 0, clip=true]{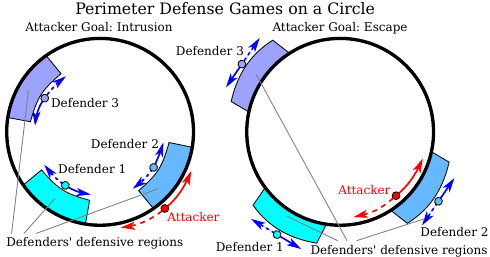}

\vspace{.2cm}

\includegraphics[width=3.4in, trim=0 0 0 0, clip=true]{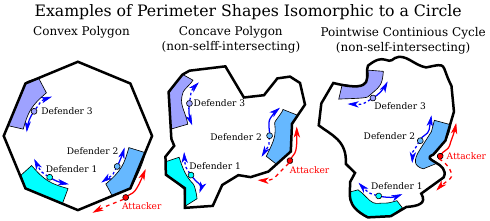}

  \caption[]{Top: The cylindrical game we consider can be used as an approximation to a defense game with a circular boundary whenever the difference in radii between inside and outside the circular boundary are negligible.
  Bottom: Circles are homomorphic to non-self-intersecting cycles of other shapes. Defense of a circular perimeter has previously been used as an approximation to defense of other boundary shapes (this requires that the effects of turning and curvature are negligible, for example, with respect to how the agents' velocities and defensive radii interact with the boundary shape). 
  }
  \label{fig:approximations}
\end{figure}

Boundary defense on a topological cylinder is a useful scenario to consider because it represents the limiting case of other perimeter defense scenarios. For example, cases of circular perimeter defense approach that of a cylinder whenever the difference in circumnavigation distance experienced by defenders and attackers approaches $0$ (see Figure~\ref{fig:approximations}, top). Previous work has remarked that continuous $1$-D boundaries without self intersections are homeomorphic to a circle \cite{Agmon.etal.JAIR11}. Thus, our results also provide insight into the patrol of other $1$-D boundaries (see Figure~\ref{fig:approximations}, bottom).  The expression derived in this paper can be used directly for other $1$-D boundaries whenever the complications of movement around corners or with curvature can be neglected.

The special case in which the attacker starts infinitesimally close to the boundary is of theoretical interest because it allows us to consider the relative trade-offs between the geometric, velocity, and defender count aspects of the game --- without the added complications of needing to consider when and how the defenders first detect the attacker. That said, the results we derive for the case where the attacker starts close to the boundary extend directly to otherwise identical cases where the attacker starts any finite distance above the boundary --- because in order to get closer to the boundary the attacker would need to turn toward the boundary, thereby decreasing its horizontal speed until it reached the boundary. 
%

Note that cases in which the attacker starts at an unguarded portion of the boundary are trivial (the attacker wins). Assuming that the attacker starts at a guarded position of the boundary, then scenarios in which the defenders move faster than the attacker are also trivial (the defenders win). In this paper we consider only the non-trivial cases: the attacker starts at a guarded region and moves faster than the defenders.

\section{Related Work}

The scenario that we consider is a multi-agent single-defender perimeter defense game. As such, it is related to other perimeter defense games, perimeter patrol games, and security games more broadly. 
It is a also a dynamic game due to the consideration of agent velocities.

The term ``security game'' has often been used to describe a scenario in which one or more defenders must protect one or more targets from one or more attackers \cite{Gatti.elal.ECAI08}. 
Many different formulations exist including:   
games defined over graphs \cite{Basilico.etal.SCIG09,Machado.etal.MASABS02,Almeida.etal.BSAI04,Tsai.etal.AAAI10,bovsansky.etal.AAMAS11,Root.PhD14},
continuous space 
\cite{Agmon.etal.JAIR11,Almeida.etal.BSAI04,Elmaliach.etal.PIJCAAMS08,Root.PhD14,GuerreroBonilla.etal.MRS21},
games considering moving targets \cite{Basilico.Gatti.AI11},
and game theoretic formulations that consider the allocation of defenders to targets without considering movement \cite{Gatti.elal.ECAI08,Kiekintveld.etal.AAMAS09,paruchuri.etal.AAMAS08}.

Perimeter patrol games can be considered a subclass of security games in which the target takes the form of an interface, perimiter, or boundary that the attacker attempts to access/cross. 
Typical formulations of perimeter patrol games assume that the defenders have limited {\it a priori} knowledge of when or where an attacker will act, and so must formulate a patrol strategy to minimize the damage that the attacker can cause \cite{Agmon.etal.JAIR11,Root.PhD14}. Many patrol games are Stackelberg games \cite{VonStackelberg.book34,%
paruchuri.etal.AAMAS08,Kiekintveld.etal.AAMAS09} such that the attacker is assumed to know the strategy that the defender(s) will use (typically, the defender team is the Stackelberg leader and the attacker is the Stackelberg follower). In the latter, it is often advantageous for the defenders to play a randomized or stochastic strategy 
so that the attacker cannot deduce when the defenders will be absent from a particular location \cite{Agmon.etal.JAIR11,Root.PhD14}.

The scenario we consider is slightly different than the aforementioned general patrol games because we assume that defenders start with knowledge of the attacker's location. The term ``perimeter defense game'' has been used in previous work \cite{Elmaliach.etal.PIJCAAMS08,GuerreroBonilla.etal.MRS21,Shishika.etal.CDC19,Shishika.etal.RAL20} to highlight this difference versus other patrol games.
One difference between our work and prior work on perimeter defense games is our assumption that the attacker starts near the perimeter's boundary. Another difference is our assumption that defenders start in an optimal configuration (though, if the attacker started further from the boundary, then the defenders would not necessarily need to be optimally distributed at time 0). 

Differential games consider players that move through space according to some prescribed dynamics of motion \cite{Isaacs.book.99,Chung.etal.AR11,Bakolas.Tsiotras.CDC10,Garcia.etal.TAC18,Shishika.etal.RAL20}. 
Prior work on perimeter patrol and/or perimeter defense games that consider agent velocities include \cite{Elmaliach.etal.PIJCAAMS08,GuerreroBonilla.etal.MRS21,Shishika.etal.CDC19,Shishika.etal.RAL20}. Reach-Avoid games \cite{Chen.etal.CDC14} are another closely related differential security game in which the defender(s) try to prevent the attacker(s) from reaching the target(s) or boundary while also considering dynamics such as velocity.

\section{Nomenclature}

\begin{figure*}[!]

\begin{minipage}{3.5in}
\begin{xy}
\xyimport(100,100){\includegraphics[width=3.4in, trim=0 0 0 0, clip=true]{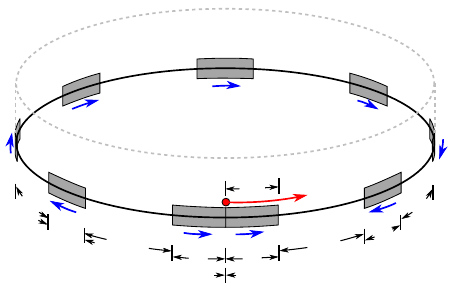}}
      ,(50,100)*{\text{Case 1}}
      ,(55.5362,33.7342)*{\text{\scriptsize$\dist_{\adversary,1}$}}
      ,(55.5362,9.266)*{\text{\scriptsize$\dist_{1}$}}
      ,(71.397,14.3192)*{\text{\scriptsize$\dist_{1,\!2}$}}
      ,(84.4195,18.3086)*{\text{\scriptsize$\dist_{2}$}}
      ,(92.4334,28.681)*{\text{\scriptsize$\dist_{2,\!3}$}}
      ,(43.6823,9.266)*{\text{\scriptsize$\dist_{\numAgents}$}}
      ,(28.1555,14.5852)*{\text{\scriptsize$\dist_{\numAgents\!-\!1,\!\numAgents}$}}
      ,(13.7981,18.5746)*{\text{\scriptsize$\dist_{\numAgents\!-\!1}$}}
      ,(5.6182,28.681)*{\text{\scriptsize$\dist_{\numAgents\!-\!2,\!\numAgents\!-\!1}$}}
      ,(70.2283,32.6704)*{\text{\scriptsize$\velocity_{\adversary}$}}
      ,(60.5449,18.3086)*{\text{\scriptsize$\velocity_{1}$}}
      ,(79.7448,23.8938)*{\text{\scriptsize$\velocity_{2}$}}
      ,(97.1082,41.181)*{\text{\scriptsize$\velocity_{3}$}}
      ,(20.2961,24.237)*{\text{\scriptsize$\velocity_{\numAgents\!-\!1}$}}
      ,(38.5067,17.7766)*{\text{\scriptsize$\velocity_{\numAgents}$}}
      ,(49.8597,-1.6382)*{\text{\scriptsize$\dist_{\numAgents,\!1}$}}
\end{xy}
\end{minipage}
\hspace{.2cm}
\begin{minipage}{3.5in}

\vspace{-.55cm}
\begin{xy}
\xyimport(100,100){\includegraphics[width=3.4in, trim=0 0 0 0, clip=true]{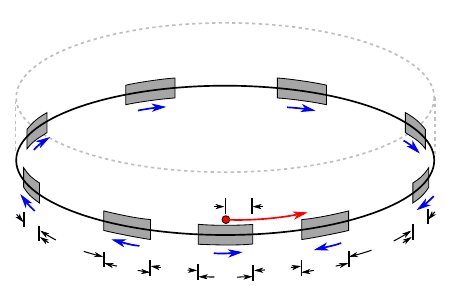}}
      ,(50,95.5852)*{\text{Case 2}}
      ,(52.6979,27.6172)*{\text{\scriptsize$\dist_{\adversary,1}$}}
      ,(49.8597,2.6172)*{\text{\scriptsize$\dist_{1}$}}
      ,(61.2127,5.2766)*{\text{\scriptsize$\dist_{1,\!2}$}}
      ,(71.8978,5.8086)*{\text{\scriptsize$\dist_{2}$}}
      ,(84.0856,14.1874)*{\text{\scriptsize$\dist_{2,\!3}$}}
      ,(93.1012,19.3724)*{\text{\scriptsize$\dist_{3}$}}
      ,(38.5067,5.5426)*{\text{\scriptsize$\dist_{\numAgents,\!1}$}}
      ,(27.8216,5.8086)*{\text{\scriptsize$\dist_{\numAgents}$}}
      ,(18.2999,14.7874)*{\text{\scriptsize$\dist_{\numAgents\!-\!2,\!\numAgents\!-\!1}$}}
      ,(4.9521,19.1064)*{\text{\scriptsize$\dist_{\numAgents\!-\!1}$}}
      ,(67.0561,29.4788)*{\text{\scriptsize$\velocity_{\adversary}$}}
      ,(55.3692,11.3938)*{\text{\scriptsize$\velocity_{1}$}}
      ,(66.8892,11.3938)*{\text{\scriptsize$\velocity_{2}$}}
      ,(90.096,24.9576)*{\text{\scriptsize$\velocity_{3}$}}
      ,(33.3919,13.271)*{\text{\scriptsize$\velocity_{\numAgents}$}}
      ,(11.6258,25.0204)*{\text{\scriptsize$\velocity_{\numAgents\!-\!1}$}}
\end{xy}
\end{minipage}

\vspace{-.1cm}

  \caption[]{Two starting scenarios for a well balanced cylinder patrol game. The attacker is red, $\numAgents$ defenders patrol a perimeter (black). The distance sensed by each defender is gray. The attacker moves with a quicker speed $\velocity_{\adversary}$ (red arrow) than the defenders' speeds $\velocity_{\agentIndex}$ (blue arrows). In the homogeneous case, all defenders move at the same speed $\velocity_{\agent} = \velocity_{\agentIndex}$ for all $\agentIndex$ and all defenders are able to sense the same distance $\dist_{\agent} = \dist_{\agentIndex}$ for all $\agentIndex$.  The length of the gap between the regions sensed by defenders $\agentIndex$ and $\agentIndexB$ is denoted $\dist_{\agentIndex,\agentIndexB}$.}
  \label{fig:idealCase}
\end{figure*}

A team of $\numAgents$ defenders, labeled $1, 2, \ldots \numAgents$, guard a perimeter located at a particular height of a cylinder. The circumference of the perimeter (and cylinder) is $\Cicumfrence$.
Without loss of generality defenders are labeled based on their starting position around the perimeter. The instantaneous location of defender $\agentIndex$ is $\pose_\agentIndex$, and defender $\agentIndex$ is able to defend a swath of distance of length $\dist_{\agentIndex}$, to travel at velocity $\velocity_\agentIndex$, and may instantaneously change direction to travel at velocity $-\velocity_\agentIndex$. 
There is a single attacker labeled $\adversary$. The instantaneous location of the attacker is $\pose_\adversary$. The attacker travels at velocity $\velocity_\adversary$ and may instantaneously change direction to travel at velocity $-\velocity_\adversary$. 

The region of the boundary that is defended by a given defender can be represented as a continuous closed interval.
As defenders move around the perimeter, the gaps between consecutive defenders may change in size. Let the length of the instantaneous gap between the regions defended by consecutive defenders $\agentIndex$ and $\agentIndexB$ be denoted $\dist_{\agentIndex,\agentIndexB}$. 
We assume that defenders coordinate such that the Lebesgue measure of the intersection of adjacent defended region is $0$. For example, if defenders $\agentIndex$ and $\agentIndex +1$ defend the adjacent continuous intervals $[0, 1]$ and $[1, 2]$ then overlap only occurs at the single point $\pose = 1$. 

For the purposes of analysis we assume that $\velocity_\adversary > \velocity_\agentIndex$ for all $\agentIndex \in \{1, \numAgents\}$ and that the game begins with the attacker located within a defended area (otherwise the game has a trivial solution).  
Without loss of generality, assume that defender $1$ is blocking the attacker. Let $\dist_{\adversary,1}$ represent 
the length of the segment defended by defender $1$ in the direction of travel of the attacker.  We assume defenders can react instantaneously to changes in the attacker's direction. For analysis (and without loss of generality) we assume they are moving in the positive direction. 

\section{Ideal Case of Homogeneous Defenders}

We now consider the case of homogeneous defenders where $\dist_{1} = \dist_{\agentIndex}$ for all $\agentIndex$
and
$\velocity_{1} = \velocity_{\agentIndex}$ for all $\agentIndex$.
For notational convenience we let $\dist_{\agent} = \dist_{\agentIndex}$ for all $\agentIndex$ 
and 
$\velocity_{\agent} = \velocity_{\agentIndex}$ for all $\agentIndex$. The optimal strategy involves the current blocking defender moving in the same direction as the attacker. This is as the attacker will slowly outrun the defender, and all other defenders must move in the reverse direction around the cylinder so that they can reinforce the perimeter ahead of the attacker.

The amount of time that defender $1$ can block the attacker, before the attacker moves beyond defender $1$'s blocking ability, is 
$
\timeVar = \frac{\dist_{\adversary,1}}{\velocity_\adversary - \velocity_\agentIndex}.
$ 

The next defender, defender $2$, must position itself so that the region it defends touches with that of agent $1$ before (or exactly at) the instant the attacker moves beyond the area defended by agent $\agentIndex$. This is only possible if 
$
\timeVar \geq \frac{\dist_{1,2}}{\velocity_1 + \velocity_2}.
$

Two extreme configurations are shown in Figure~\ref{fig:idealCase} left and right, respectively. 
\begin{enumerate}
\item The attacker is at the single point where the regions defended by defenders $\numAgents$ and $1$ touch. 
\item The attacker is at the midpoint of the area defended by defender $1$.
\end{enumerate}
In either case, the defenders are positioned exactly such that, if the attacker moves right (or left, respectively) across the distanced defended by defender $\agentIndex$ then the next defender $\agentIndex + 1$  (or $\agentIndex - 1$, respectively) moves toward the attacker so that it can begin blocking the attacker the instant the attacker reaches the end of defender $\agentIndex$'s blocking abilities.

We shall calculate the maximum circumference $\Cicumfrence$ that $n$ defenders can successfully defend given either starting condition, and show that the two are identical.

\noindent {\bf Case 1:} $\dist_{\adversary,1} = \dist_{1} = \dist_{\agent}$ and $\dist_{\numAgents,0} = 0$. Due to symmetry $\dist_{1,2} = \dist_{\agentIndex,\agentIndex+1}$ for all $\agentIndex$. In other words, all gap lengths between defended regions are equal in this case. For notational convenience we denote $\dist_{\gap} = \dist_{\agentIndex,\agentIndex+1}$. The circumference is found by summing over defended regions and gaps between them.

$\displaystyle
\Cicumfrence_{max} = \sum_{\agentIndex = 1}^{\numAgents} \dist_{\agentIndex} + \sum_{\agentIndex = 1}^{\numAgents-1} \dist_{\agentIndex,\agentIndex+1} +  \dist_{\numAgents,0} = \numAgents \dist_{\agent} + (\numAgents-1)  \dist_{\gap}
$

\noindent {\bf Case 2:} In this case the attacker starts halfway along the distance defended by defender $1$ such that $\dist_{\adversary,1} = \frac{\dist_{1}}{2}=\frac{\dist_{\agent}}{2}$. Due to symmetry, the gaps between regions defended by defender $2$ through $\numAgents$ is the same as in Case 1 (note, however, that the the gap between the regions defended by defenders $1$ and $2$ is not the same).
$\dist_{2,3} = \dist_{\agentIndex,\agentIndex+1} = \dist_{\gap}$ for all $\agentIndex \geq 2$.

Given the assumption of constant velocity as well as the construction that defender $2$ starts to defend the attacker the last instant it is defended by defender $1$, the maximum distance between the regions defended by defender $1$ and $2$ must be half of $\dist_{\gap}$. Formally,  $\dist_{1,2} = \frac{1}{2} \dist_{\gap}$.
Due to symmetry, the gap between regions defended by defenders $\numAgents$ and $1$ is equal to the gap between regions defended by defenders $1$ and $2$. So, $\dist_{\numAgents,1} = \dist_{1,2} = \frac{1}{2} \dist_{\gap}$.
Again, the circumference is found by summing up defended regions and the gaps between them.
\begin{align*}
\Cicumfrence_{max} &= \sum_{\agentIndex = 1}^{\numAgents} \dist_{\agentIndex} + \sum_{\agentIndex = 1}^{\numAgents-1} \dist_{\agentIndex,\agentIndex+1} +  \dist_{\numAgents,0} \\
                   & = \numAgents \dist_{\agent} +  \frac{1}{2} \dist_{\gap} + (\numAgents-2)  \dist_{\gap} + \frac{1}{2} \dist_{\gap} \\
                   & = \numAgents \dist_{\agent} + (\numAgents-1)  \dist_{\gap}
\end{align*}
This result is identical to that for Case 1. The fact that the circumferences resulting from both cases are the same can be seen by noting the following: after the attacker moves into the area defended by defender $1$ all other defender (except defender $1$) move at the same speed in the opposite direction as defender $1$. Thus, length is added to $\dist_{\numAgents,1}$ at the same rate it is removed from  $\dist_{1,2}$.  Extending this line of reasoning it is apparent that cases 1 and 2 represent the same scenario at different times. In particular, due to constant velocities, Case 2 happens a duration of time after Case 1 given by
$$
\timeVar_{\mathrm{1,2}} = \frac{\timeVar}{2} = \left(\frac{1}{2}\right)\frac{\dist_{\gap}}{2\velocity_{\agent}} = \left(\frac{1}{2}\right)\frac{\dist_{\agent}}{\velocity_{\adversary} - \velocity_{\agent}}.
$$

\noindent {\bf Optimal gap distances, derived from Case 1:}

With the exception of $\dist_{n,1}$ and $\dist_{1,2}$ all gap distance remain the same in Cases 1 and Case 2. Focusing on the situation in Case 1, $\dist_{1,2} = \ldots = \dist_{n-1,2} = \dist_{\gap}$ can be calculated as a function of defender and attacker velocities and  $\dist_{\agent}$ by solving for the maximum gap distance that two defenders moving toward each other can close in time $\timeVar$. Without loss of generality, we consider defenders i=1 and i+1=2 (considering Case 1):

\begin{minipage}{\columnwidth}
\centering 
\vspace{.2cm}

$\displaystyle
\frac{\dist_{\agent}}{\velocity_\adversary - \velocity_{\agent}}
=
\frac{\dist_{\adversary,1}}{\velocity_\adversary - \velocity_1} 
=
\timeVar
=
\frac{\dist_{1,2}}{\velocity_1 + \velocity_2}
=
\frac{\dist_{\gap}}{2 \velocity_{\agent}}
$

\vspace{.2cm}

$\displaystyle 
\dist_{\gap}
=
\dist_{\agent} \frac{2 \velocity_{\agent}}{\velocity_\adversary - \velocity_{\agent}}
$
\end{minipage}

\vspace{.2cm}

\hspace{0.5cm}
$\displaystyle
\Cicumfrence_{max} = \numAgents \dist_{\agent} + (\numAgents-1) \dist_{\agent} \frac{2 \velocity_{\agent}}{\velocity_\adversary - \velocity_{\agent}}.
$
\vspace{0.5cm}

\noindent All else being equal, the attacker will win if $\Cicumfrence > \Cicumfrence_{max}$ because there will eventually be a nonzero gap between the regions defended by consecutive defenders. The attacker will win if:

\hspace{1cm}$\displaystyle
\Cicumfrence > \numAgents \dist_{\agent} + (\numAgents-1) \dist_{\agent} \frac{2 \velocity_{\agent}}{\velocity_\adversary - \velocity_{\agent}}.
$
\vspace{0.2cm}

\noindent This inequality can also be rearranged to highlight the bounding constraints on $\numAgents$, $\velocity_{\agent}$, and $\velocity_\adversary$ in terms of the other quantities and $\Cicumfrence$. Let 
$
\param = \frac{2 \velocity_{\agent}}{\velocity_\adversary - \velocity_{\agent}}.
$
The attacker will win if:
$$
\frac{\Cicumfrence}{\dist_{\agent}} > \numAgents  + (\numAgents-1) \param
\hspace{.4cm}
\text{ or } 
\hspace{.4cm}
\dist_{\agent} \! < \! \frac{\Cicumfrence}{\numAgents  + (\numAgents-1) \param}
$$
$$
\text{or }
\hspace{.2cm}
\param \! < \! \frac{(\Cicumfrence / \dist_{\agent}) \! - \! \numAgents}{\numAgents \! - \!1}
\hspace{.2cm}
\text{ or }
\hspace{.2cm}
\frac{\velocity_\adversary}{\velocity_{\agent}} \! >  \! \frac{2(\numAgents \! - \! 1)}{(\Cicumfrence / \dist_{\agent}) \! - \! \numAgents} \! + \! 1.
$$

\section{Summary and Conclusions}

This paper has considered the special case of the multi-defender single-attacker perimeter defense game on a cylinder. The special case that we consider is that the attacker starts infinitesimally close to a portion of the boundary that is currently guarded by a defender and defenders are optimally positioned, yet the attacker is able to move more quickly than any of the (homogeneous) defenders.  Assuming that the defending team is able to coordinate their actions, we have derived closed-form expression describing when the attacker will win. 

The expression that we derive are potentially useful for quantifying the relative trade-offs between having more defenders, faster defenders, and/or defenders with a greater defensive range.

\bibliographystyle{IEEEtran}
\bibliography{bibliography}

\end{document}